\documentclass[%
 reprint,
 amsmath,amssymb,
 aps,
 prapplied,
 floatfix,
 superscriptaddress,
 longbibliography,
]{revtex4-2}
\usepackage{comment}

\usepackage[english]{babel} 
\usepackage{xcolor}
\usepackage{graphicx} 
\usepackage{dcolumn} 
\usepackage{bm} 
\usepackage[bb=boondox]{mathalfa}
\usepackage{amsmath}
\usepackage{stmaryrd}

\usepackage{siunitx}
\usepackage{tabularx}
\usepackage{xr}
\usepackage{hyperref} 

\makeatletter
\newcommand*{\addFileDependency}[1]{
\typeout{(#1)}
\@addtofilelist{#1}
\IfFileExists{#1}{}{\typeout{No file #1.}}
}\makeatother
\usepackage[indention = 0pt]{caption}
\usepackage{subcaption}
\begin{document}

\title{Experimental quantum reservoir computing with a circuit quantum electrodynamics system}

\author{B.~Carles}
\author{J.~Dudas}
\affiliation{
Laboratoire Albert Fert, CNRS, Thales, Université Paris-Saclay, 91767 Palaiseau, France
}
\author{L. Balembois}
\affiliation{SPEC, CEA, Gif-sur-Yvette, France }

\author{J. Grollier}
\author{D.~Marković}
\email{Corresponding author: danijela.markovic@cnrs.fr}
\affiliation{
Laboratoire Albert Fert, CNRS, Thales, Université Paris-Saclay, 91767 Palaiseau, France
}

\date{\today}

\begin{abstract}

Quantum reservoir computing is a machine learning framework that offers ease of training compared to other quantum neural networks, as it does not rely on gradient-based optimization. Learning is performed in a single step on the output features measured from the quantum system. Various implementations of quantum reservoir computing have been explored in simulations, with different measured features. Although simulations have shown that quantum reservoirs present advantages in performance compared to classical reservoirs, experimental implementations have remained scarce. This is due to the challenge of obtaining a large number of output features that are nonlinear transformations of the input data. In this work, we show that even with a circuit quantum electrodynamics system as simple as a single transmon coupled to a readout resonator, we can implement a
proof-of-concept realization of quantum reservoir computing. We obtain a large number of nonlinear features from a single physical system by encoding the input data in the amplitude of a coherent drive and measuring the cavity state in the Fock basis. We demonstrate classification of two classical tasks with significantly smaller hardware resources and fewer measured features compared to classical neural networks. Our experimental results are supported by numerical simulations that show additional Kerr nonlinearity is beneficial to reservoir performance. Our work demonstrates a hardware-efficient quantum neural network implementation that can be further scaled up and generalized to other quantum machine learning models.

\end{abstract}

\maketitle

\section{Introduction}

Quantum reservoir computing was introduced as an approach to quantum machine learning that circumvents the need to train quantum systems directly. This is particularly appealing because gradient-based optimization methods that work so well for classical artificial neural networks are difficult to apply to quantum systems: estimating gradients is experimentally demanding, and the optimization landscape often suffers from barren plateaus that hinder convergence~\cite{larocca_barren_2025}.
Quantum reservoir computing has been explored in simulations using a variety of systems, including qubit-based reservoirs~\cite{fujii_harnessing_2017, ghosh_quantum_2019} and bosonic modes~\cite{govia_quantum_2021, nokkala_gaussian_2021, angelatos_reservoir_2021}. These studies have shown that QRC can solve nontrivial tasks with fewer output neurons than classical reservoirs, highlighting both the computational advantage provided by quantum coherence and the potential for processing quantum input data~\cite{ghosh_quantum_2019}. Experimental realizations of quantum reservoir computing, however, remain scarce. Existing implementations include demonstrations on qubit-based quantum processors available through online platforms~\cite{hu_overcoming_2024, suzuki_natural_2022, chen_temporal_2020}, quantum extreme learning machines with nuclear spins~\cite{negoro_machine_2018} and with photon orbital angular momentum~\cite{suprano_experimental_2024, zia_quantum_2025}  and, more recently, an analog realization using a bosonic mode coupled to a qubit~\cite{senanian_microwave_2024}. Here, we report on a hardware-efficient implementation of quantum reservoir computing using a single bosonic mode hosted by a superconducting resonator, coupled to an ancilla qubit for readout. In our scheme, the input data are encoded in the amplitude of the cavity displacement, while the reservoir outputs are given by the Fock-state occupation probabilities. This approach provides a large number of nonlinear features directly from the quantum dynamics. Importantly, the nonlinearity arises from the measurement process itself rather than from the input encoding, a key property for future extensions toward learning on quantum data.

\begin{figure}[h!]
    \centering
    \includegraphics[width=\linewidth]{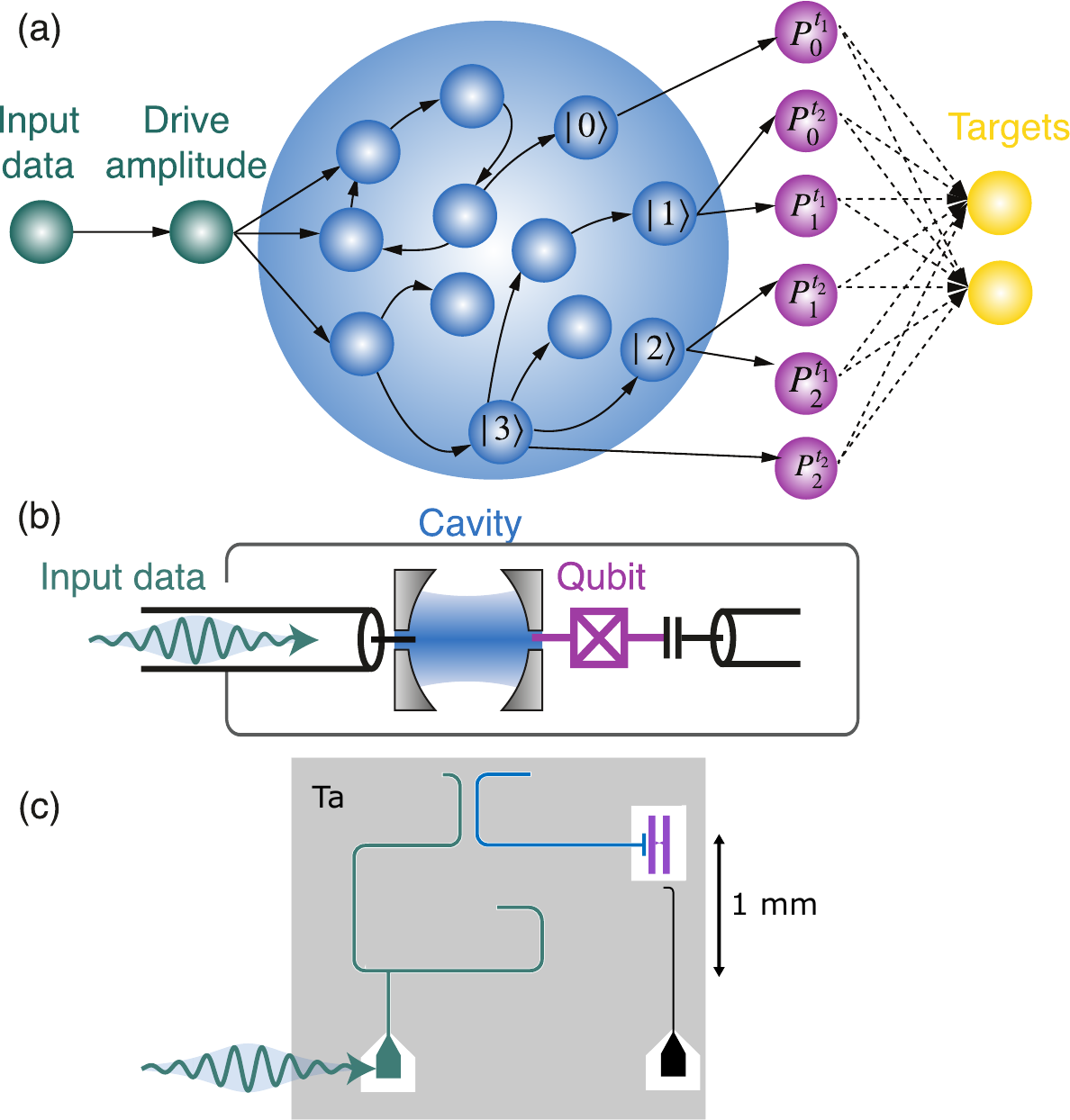}
    \caption{\label{Figure1} \justifying \bf{Quantum reservoir computing with a superconducting circuit.} \rm
    (a) Input data are encoded in the amplitude of the resonant drive (green) of the quantum system acting as a reservoir (blue). Output features (purple circles) are obtained by measuring the occupation probabilities of Fock states $|0\rangle$ to $|n\rangle$ at different times $t_1$ and $t_2$, yielding a total of $2n$ output features. Features are classified by training a layer of linear weights (dashed lines). (b) Schematic and (c) design of the quantum circuit used to implement the quantum reservoir. Tantalum coplanar waveguide resonator (blue) is capacitively coupled to a transmon qubit (purple) and to a transmission line through a Purcell filter (green).}
\end{figure}

\section{Circuit and implementation}

\begin{figure}[htp]
    \centering
    \includegraphics[width=\linewidth]{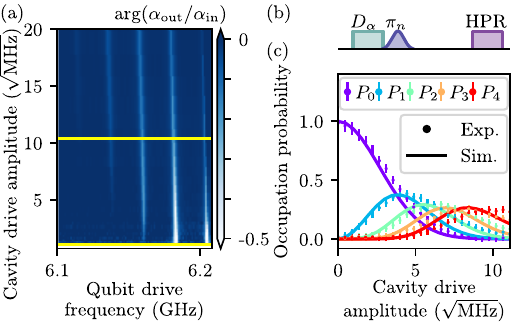}
    \caption{\label{Figure2} \justifying \bf Fock state probabilities as neurons. \rm (a) Two tone spectroscopy of the cavity. The phase shift of the reflection coefficient at the cavity resonance frequency, as a function of the qubit drive frequency and the cavity drive amplitude $\alpha_{\rm{in}}(t)$. Different resonances from right to left correspond to the qubit frequency dressed by 0, 1, 2, 3 and 4 photons in the cavity. Yellow horizontal lines indicate the amplitude range used for data encoding. (b) Top: pulse sequence used to measure the photon number in the cavity. A 200~ns displacement pulse $D_{\alpha}$ at cavity frequency $\omega_c$ is followed by a 200~ns $\pi_n$ pulse conditioned on $n$ photons in the cavity, and a high power readout (HPR) after the 1~$\mu$s waiting time. Bottom: Measured occupation probabilities (dots) $P_0$ to $P_4$ for Fock states $|0\rangle$ to $|4\rangle$ as a function of the cavity displacement amplitude $\alpha_{\rm{in}}$. Each point is averaged 10 000 times. Lines correspond to the probabilities obtained from the simulations using the Lindbladt master equation, with dissipation rates and Kerr effect as free parameters.} 
\end{figure}

We implement quantum reservoir computing on a superconducting circuit composed of a Tantalum coplanar waveguide resonator, capacitively coupled to a transmon qubit (Figure~\ref{Figure1}(b) and (c)). We use the fundamental $\lambda/2$ mode of the resonator as a cavity, at a frequency $\omega_c = 2 \pi \times 7.617$ GHz. We use the transmon qubit as ancilla, to measure the occupation probabilities of the cavity Fock states that serve the function of output neurons (Figure~\ref{Figure1}(a)). The cavity-qubit system can be described by the Hamiltonian
\begin{equation}
    \frac{\hat{H}}{\hbar} = \omega_c \hat{a}^\dagger \hat{a} +   \omega_q \frac{\hat{\sigma}_Z}{2} -  \frac{\chi}{2} \hat{a}^\dagger \hat{a} \hat{\sigma}_Z + K_{cc} (\hat{a}^\dagger\hat{a})^2 +K_{cq} (\hat{a}^\dagger\hat{a})^2 \frac{\hat{\sigma}_Z}{2},
    \label{hamiltonien}
\end{equation}
where $\omega_q = 2 \pi \times 6.210$~GHz is the bare qubit frequency, $\chi = 2 \pi \times 22.29$~MHz is the dispersive coupling rate, $K_{cc} = -2 \pi \times 300$ kHz is the cavity self-Kerr coefficient and  $K_{cq}\approx - 2\pi \times 0.44$ MHz is a photon-number–dependent correction to the dispersive shift. The dispersive coupling makes the qubit resonance frequency depend on the number of photons $n= \langle \hat{a}^\dagger \hat{a} \rangle$ in the cavity. The cavity drive Hamiltonian is
\begin{equation}
    \hat{H}_{\rm{drive}} = \hbar \sqrt{\kappa_{\rm{ext}}} \alpha_{\rm{in}}(t) ( \hat{a} + \hat{a}^\dagger ),
\end{equation}
where $\kappa_{\rm{ext}}$ is the cavity coupling to the transmission line and $\alpha_{\rm{in}}(t)$ is the amplitude of the classical cavity drive. In the two tone spectroscopy shown in Figure~\ref{Figure2}(a), we observe a set of equally spaced drive frequencies that lead to a significant shift of the cavity resonance. These drive frequencies, separated by $\chi$, correspond to resonance frequencies $\omega_q^n = \omega_q - n \chi$ of the qubit dressed by $n$ photons in the cavity. As is commonly done in circuit QED~\cite{schuster_resolving_2007}, we use these photon number dependent resonances to measure the number of photons in the cavity, and to infer the Fock state occupation probabilities from the measurement statistics. The time sequence used to measure the occupation probabilities is shown in the top of Figure~\ref{Figure2}(b). A 200 ns cavity displacement pulse $D_{\alpha}$ is applied at cavity resonance associated with the transmon being in the ground state, followed by a photon number-conditioned $\pi_n$ pulse applied to the qubit at frequency $\omega_q^n$. The state of the qubit is read through the same resonator, using high-power readout~\cite{reed_high-fidelity_2010}. For this reason, we perform the readout 1 $\mu$s after the $\pi_n$ pulse, letting enough time for the resonator to empty, such that the readout pulse is not affected by the photons residual from the input data. In future experiments, this delay could be reduced by employing a dedicated readout resonator coupled to the qubit. Furthermore, the protocol could be made more efficient by using the multiplexed readout~\cite{heinsoo_rapid_2018, essig_multiplexed_2021} and deep neural network discrimination~\cite{lienhard_deep-neural-network_2022}.

\section{Characterization of neural nonlinearities}

Fock state populations of a coherent state $|\alpha\rangle$ are given by Poissonian distributions, $P_n(\alpha) = e^{-|\alpha|^2} \frac{|\alpha|^{2n}}{n!}$, which have a nonlinear dependence on the cavity coherent state amplitude $\alpha$, that can be exploited as a neural activation function. In our circuit, the measured distributions deviate from Poisson statistics as a result of the Kerr effect originating from the nonlinearity inherited by the cavity from the qubit. We simulate the probabilities using the Lindblad master equation, with the dissipation rates and cavity self-Kerr coefficient as free parameters determined by fitting to the measured data (Figure \ref{Figure2}(b)). The extracted values, $\kappa_{\rm{tot}} = \kappa_{\rm{ext}} + \kappa_{\rm{int}} = 2 \pi \times (560 \pm 60)$~kHz and $K_{\rm{cc}} = 2 \pi \times (-300 \pm 50)$~kHz, are consistent with those obtained from reflection measurements.

Due to the self-Kerr effect, the cavity mode frequency shifts at high input amplitudes corresponding to large data values. Within the encoding range $\alpha_{\rm{in}} \in [1, 10.4] \sqrt{\mathrm{MHz}}$, the cavity resonance shifts by $\Delta f_c = 1.7$~MHz, which largely exceeds the cavity linewidth $\kappa_{\rm{tot}}$. To account for this shift, we use short input pulses of duration $\Delta t_{\rm{in}} = 200$~ns, giving a spectral width $\Delta t_{\rm{in}}^{-1} = 5~\rm{MHz} > \Delta f_c$. Furthermore, a higher order correction, corresponding to a photon number dependent dispersive shift $K_{cq}$, reduces the efficiency of the conditional $\pi_n$ pulse and, consequently, the overall readout efficiency.

 \begin{figure*}[htp]
    \centering
    \includegraphics[width=\linewidth]{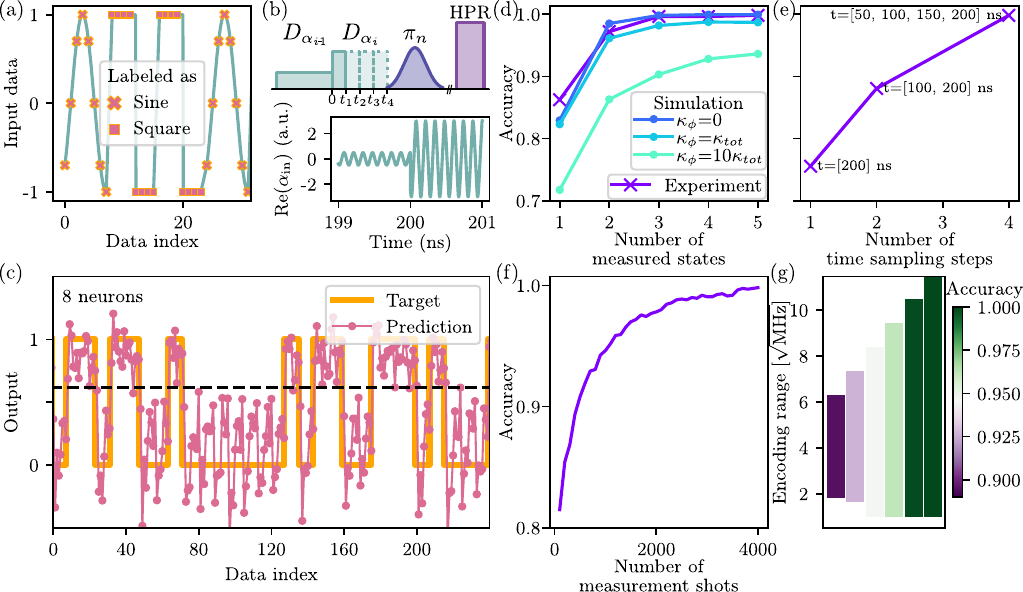}
    \caption{\label{Figure3} \bf{Sine and square waveform classification task.} \rm \justifying (a) Input data are obtained from a random series of 400 sine and square periods, discretized into 8 points per period. Only 4 periods are shown for clarity. (b) Top: Pulse sequence consists of two displacement pulses $D_{\alpha_{i-1}}$ and $D_{\alpha_{i}}$, whose amplitudes encode two consecutive data points, followed by a $\pi_{n}$ pulse and high-power readout. To increase the number of extracted features, the occupation probabilities are sampled at four times $t_1$ to $t_4$ during the second displacement pulse. Bottom: Cavity drive $\mathrm{Re}(\alpha_\mathrm{in})$ whose amplitude encodes the input data. (c) Prediction of the quantum reservoir with the 8 most informative features among probabilities of states $|0\rangle$ to $|4\rangle$ at times $t_i=i\times50$~ns, $i \in [ 1, 4 ]$. (d) Experimental accuracy (purple crosses) as a function of the number of measured states $N$, with 4 time samples per state, giving $4N$ features. For each $N$, the plot shows the best accuracy obtained with the $N$ most informative features. The solid lines connect simulated data points (dots) corresponding to different qubit dephasing rates $\kappa_{\phi}$. (e) Experimental accuracy with 5 measured features (states from $|0\rangle$ to $|4\rangle$ as a function of the number of time sampling steps, with sampling times indicated in brackets). (f) Accuracy as a function of the number of measurement shots and (g) as a function of the encoding range. }
\end{figure*}

\section{Results: learning temporal tasks}

\subsection{Sine and square waveform classification}

We first deploy our quantum reservoir computing scheme on the task of sine and square waveform classification. This task consists in sending one by one, points that belong either to a sine or to a square period, and assigning them the correct label, as shown in Figure~\ref{Figure3}(a). This task probes both the nonlinearity and memory of the underlying physical system: a fully linear network can only apply a single amplitude threshold, achieving at best 68.75\% accuracy, while classical recurrent neural networks combining nonlinearity and memory can reach 100\% accuracy with 25 neurons~\cite{dudas_quantum_2023}.

We have built a dataset of 400 random sine and square periods, each discretized into 8 points. Half of the data is used for training and the remaining half for testing. The prediction
\begin{equation}
    Y = FW
\end{equation}
is obtained by multiplying the feature matrix $F$ by the weight matrix $W$. The feature matrix is composed of the occupation probabilities of states $|0\rangle$ to $|4\rangle$ measured at times $t_i = i\times 50$~ns, $i\in [ 1, 4]$ after the beginning of the pulse, and a bias term. Each probability is reconstructed from  4000 measurements.

After each measurement, the reservoir state is reinitialized by waiting 60~$\mu$s for the qubit to relax to its ground state before re-injecting the same input $x_n$ preceded by the input $x_{n-1}$, because this task requires the memory of at least one previous input. This procedure yields a total of $N_F = 20$ output features, corresponding to 5 probabilities at 4 different times. The weight matrix $W$ is computed on the training data using the ridge regression, by minimizing the cost function
 \begin{equation}
     J(W) = |Y-FW|^2 + \beta |W|^2.
 \end{equation}
 The second term is a penalty on large weights, controlled by the regularization parameter $\beta$. With 20 measured features, yielding 21 output neurons including the bias, we find that the prediction on the test set matches the target with 99.8\% accuracy. Interestingly, even when we retain only the eight most informative features, { $P_0^{t2}$, $P_0^{t3}$, $P_0^{t4}$, $P_1^{t3}$, $P_2^{t1}$, $P_2^{t2}$, $P_3^{t1}$, $P_4^{t1}$ }, we still achieve 99.5~\% accuracy, as shown in Figure~\ref{Figure3}(c). This is a significantly smaller number of features that need to be measured compared to classical reservoir computing, where at least twenty-five classical features are required to achieve comparable performance, and is consistent with previous simulations~\cite{dudas_quantum_2023}. We can notice that the most informative features correspond to a mix of different states measured at different times. In a practical implementation, the most informative features could be identified during the training phase, such that during inference only they are measured. 
 
In practice, the extraction of these features is subject to experimental imperfections that distort the measured occupation probabilities. Specifically, we measure the probability of ionizing the qubit after a $\pi_n$ pulse followed by a high-power readout pulse. Under ideal conditions, this probability would be equal to the underlying Fock-state probability $P_n$. However, due to thermal noise, finite efficiency of the $\pi_n$ pulse, finite ionization probability and qubit decay during the waiting time between the $\pi_n$ pulse and the readout, the measured switching probability $p_s(n)$ becomes an affine function of the true probability, $p_s(n) = a P_n +b$. These effects modify the output probabilities but do not degrade the performance of the quantum reservoir, since the learned weights can compensate for such distortions. Measurement imperfections nevertheless tend to concentrate  $p_s(n)$ toward 0.5, increasing the number of measurement shots required to reach a given statistical precision.

Finally, beyond such technical imperfections, we analyze the impact of decoherence on the learning performance. To pinpoint the role of quantum coherence in learning this task~\cite{palacios_role_2024}, we simulate the system for different values of the dephasing rate $\kappa_{\phi}$. We observe that the accuracy decreases significantly as dephasing increases, even when a larger number of Fock states is measured (Figure~\ref{Figure3}(d)). Figures~\ref{Figure3}(e–g) show the dependence of the experimental reservoir performance on the number of time-sampling steps, the finite number of measurement shots~\cite{khan_physical_2021}, and the amplitude encoding range. Interestingly, we find that using a larger encoding range is advantageous, despite the fact that at higher drive amplitudes the cavity population saturates due to the self-Kerr effect, and the $\pi_n$ pulses become less efficient because of the cross-Kerr effect. This behavior will be further analyzed in simulations described in Section~\ref{Section-simulations}.

\subsection{Mackey-Glass chaotic time series prediction}

\begin{figure}[htp]
    \centering
    \includegraphics[width=\linewidth]{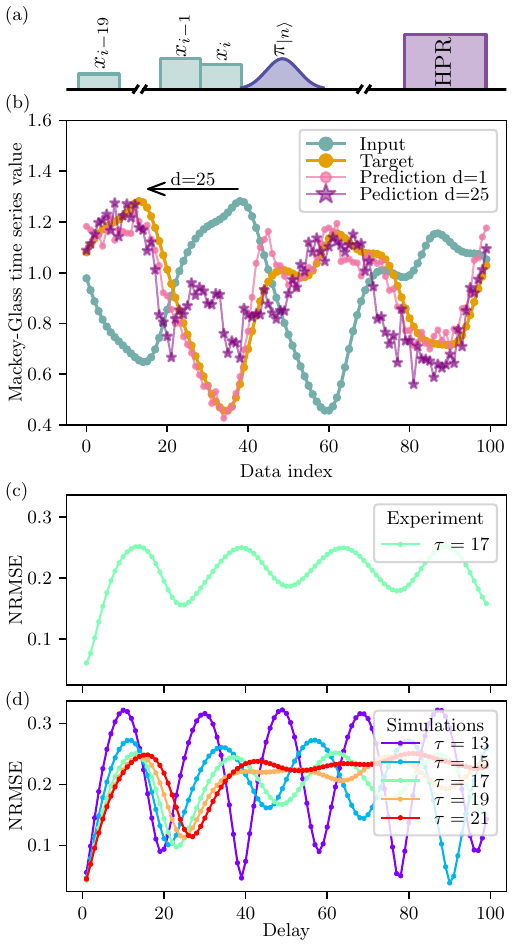}
    \caption{\label{FigureMG} \justifying \bf Mackey-Glass chaotic time series prediction. \rm (a) Pulse sequence used for the task. (b) Part of the input data (green) and corresponding target for a delay $d=25$ (gold). Quantum reservoir predictions for delays $d=1$ and $d=25$ are shown respectively in pink and purple. (c) Normalized root mean square error (NRMSE) as a function of delay $d$ for the experimental quantum reservoir for the lag parameter $\tau = 17$.~(d)~NRMSE as a function of the delay $d$ for a simulated quantum reservoir for different parameters $\tau$.} 
    \label{Mackey-Glass}
\end{figure}

We implement a second benchmark task which is Mackey-Glass chaotic time series prediction~\cite{mackey-glass}. This task consists in forecasting future values of a time series based on its past values. The time series is generated as a solution of the Mackey-Glass nonlinear differential equation 
\begin{equation}
     \frac{\partial x(t)}{\partial t} = \frac{\beta x (t-\tau)}{1+x(t-\tau)^{10}} - \gamma x(t),
\end{equation}
and is chaotic for $\beta=0.2$, $\gamma=0.1$ and $\tau = 17$. The whole series contains 2000 data points. We use half of the dataset for training and the remaining half for testing. For each value of the delay $d$, the system is trained to try to output a time-delayed version of the Mackey-Glass time series. The target is thus $\tilde{y}_k = x_{k+d}$, as shown in the Figure~\ref{FigureMG}(a). The feature vector is composed of 5 occupation probabilities of states $|0\rangle$ to $|4\rangle$ measured 100~ns after the beginning of the input encoding pulse.  For each measurement of the feature vector for the input $k$, we re-inject 20 preceding inputs $[x_{k-19}, ...., x_{k}]$, each for 100~ns. We evaluate performance using the normalized root mean square error (NRMSE),
\begin{equation}
    \mathrm{NRMSE} = \frac{1}{\Delta \tilde y} \sqrt{\frac{\sum_k (y_k-\tilde y_k)^2}{N}},
\end{equation}
where the normalization is ensured by the division by the range of the target values $ \Delta \tilde y = \mathrm{max}(\tilde y) - \mathrm{min}(\tilde y)$. Figure~\ref{Mackey-Glass}(c) shows the NRMSE as a function of the prediction delay $d$ for the lag parameter $\tau = 17$, value for which the Mackey–Glass series becomes quasi-chaotic. We observe that the NRMSE exhibits minima for delays $d$ corresponding to multiples of the quasi-oscillation period, for which the target signal is highly correlated with the input. To further illustrate this effect, we show in Figure~\ref{Mackey-Glass}(d) simulated NRMSE values for Mackey–Glass time series with different delay parameters $\tau$, ranging from smaller (periodic) to larger (chaotic) regimes. For periodic inputs, clear oscillations in the NRMSE are visible, whereas for highly chaotic inputs, the NRMSE rapidly converges to an average value.

\section{Simulations of the impact of the Kerr effect}
\label{Section-simulations}

Having experimentally demonstrated the classification of sine and square waveforms, as well as the prediction of a chaotic time series using a quantum reservoir, we now turn to simulations to explore the physical origin of the reservoir’s computational capabilities. In particular, we investigate how the Kerr nonlinearity—which plays a central role in shaping the system’s dynamics—affects performance.

The effect of different types and magnitudes of nonlinearities on the performance of quantum neural networks remains an open question~\cite{stanev_deterministic_2023, motamedi_quantumness_2024}. In optical quantum neural networks, some studies have suggested that weaker Kerr nonlinearities lead to degraded performance~\cite{steinbrecher_quantum_2019}, while in the context of bosonic quantum neural networks, it has been shown that whereas simple classical tasks such as XOR gate learning can be solved with only the encoding nonlinearity, quantum tasks require an additional Kerr nonlinearity, which also improves robustness to errors and noise~\cite{xu_roles_2023}.

\begin{figure}[h!]
    \centering
    \includegraphics[width=1\linewidth]{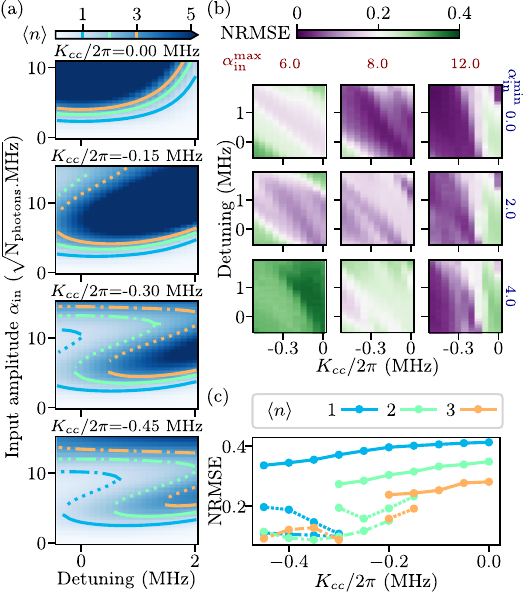}
    \caption{\label{simulations} \justifying \bf Impact of the Kerr effect on performance. \rm (a) Simulated mean photon number $\langle n \rangle$ in the cavity after a 400 ns drive as a function of detuning and input amplitude $\alpha_{\rm{in}}$, for different Kerr rates $K_{cc}$. Contour lines indicate regions of equal mean photon number. (b) NRMSE for sine-square classification using 20 neurons, shown as a function of Kerr coefficient and detuning, for various input encoding ranges [$\alpha_\mathrm{in}^\mathrm{min}$, $\alpha_\mathrm{in}^\mathrm{max}$]. 
    (c) Simulated NRMSE as a function of the Kerr coefficient $K_{cc}$ for $\Delta = 0$ and the encoding range [0.7 $\alpha_0$, 1.3 $\alpha_0$], where $\alpha_0$ is the drive amplitude required to reach $\langle n \rangle \in $ \{1, 2, 3\} after a 400~ns drive (see panel (a)). Because multiple combinations of detuning and amplitude can yield the same mean photon number, the NRMSE values for different amplitudes are shown with distinct line styles.   
    }
    
\end{figure}

To investigate the influence of Kerr nonlinearity in our setup, we simulate the classification of sine and square waveforms using a Kerr oscillator with varying Kerr coefficients $K_{cc}$. Since the input drive is applied at a fixed frequency for all drive amplitudes, its coupling to the resonator becomes less efficient when the Kerr coefficient is large. This significantly affects the mode population. Figure~\ref{simulations}(a) shows the average photon number in the mode as a function of the drive detuning from the cavity resonance at low power, $\Delta = \omega_c - \omega$, and the drive amplitude  $\alpha_{\rm{in}}$, after two 200 ns input pulses. We observe that for large Kerr coefficients $K_{cc}$ and zero detuning, the mean photon number remains below 2. Indeed, in this case, the drive is too far from the effective resonance to efficiently populate the cavity. For larger detunings, the mean photon number is a non-monotonic function of the drive amplitude: it initially increases, reaching a maximum at an intermediate amplitude where the drive frequency is closest to resonance, and then decreases again at higher amplitudes as the system moves away from resonance. This implies that for large Kerr coefficients, it will potentially be interesting to apply the input data at a frequency detuned by $\Delta \neq 0$ from the resonance.

In the following, we study the performance on the sin-square task as a function of Kerr coefficient $K_{cc}$. In order to separate the impact of the additionnal nonlinearity from the spurious effects such as the photon number in the cavity related to the frequency shift, we perform the study for different drive detunings and also for different encoding ranges $[\alpha_{\rm{min}}, \alpha_{\rm{max}}]$ as shown in Figure~\ref{simulations}(b). \noindent We observe that the lowest classification errors are achieved for the largest encoding ranges and for high values of the Kerr coefficient. 

To isolate the contribution of the Kerr effect from that of the amplitude encoding range, we proceed as follows: for a given Kerr coefficient and a fixed target mean photon number $\langle n \rangle \in \{1,2,3\}$, we select all combinations of $\{\Delta, \alpha_0 \}$ such that the lowest input amplitude results in $\langle n \rangle$ photons in the cavity. We then define the encoding range as $[0.7 \alpha_{0}, 1.3 \alpha_{0}]$. As shown in Figure~\ref{simulations}(c), we observe that the lowest errors are obtained for large Kerr coefficients. It is important to note that due to the non-monotonic behavior of the photon number as a function of input amplitude, for a given Kerr coefficient, different drive amplitudes can yield the same average photon number, as shown in Figure~\ref{simulations}(a). This explains the presence of multiple points with the same Kerr coefficient but different amplitudes in Figure~\ref{simulations}(c). Lower errors are generally associated with higher drive amplitudes. However, in practice, the maximum amplitude is limited by the onset of parametric oscillation~\cite{bengtsson_nondegenerate_2018}.

\section{Conclusion and outlook}

We have experimentally implemented quantum reservoir computing using a hardware-efficient platform composed of a single bosonic mode coupled to a qubit. By encoding input data in the cavity-mode displacement amplitudes and performing readout in the Fock basis, we extract a large number of nonlinear output features. This is essential in reservoir computing, where only the output weights are trained, and therefore a rich set of features is required. Additional features can alternatively be generated through time multiplexing~\cite{fujii_harnessing_2017}, by weak continuous monitoring of multiple observables~\cite{zhu_practical_2025}, by increasing the number of physical components in the system~\cite{hu_overcoming_2024, kornjaca_large-scale_2024}, or by computing statistical moments of the measured data~\cite{senanian_microwave_2024}. 

Nonlinear output features are necessary for successful information processing. In quantum bosonic systems, nonlinear transformations may arise from the encoding of input data~\cite{nokkala_gaussian_2021}, from intrinsic Kerr nonlinearity~\cite{govia_quantum_2021, angelatos_reservoir_2021}, or from the measurement process \cite{dudas_quantum_2023, senanian_microwave_2024}. However, for applications involving quantum input data \cite{ghosh_quantum_2019, markovic_quantum_2020, huang_quantum_2022, tran_learning_2021}, where the reservoir must interface seamlessly with a quantum system, nonlinearity cannot rely solely on the encoding stage. While Kerr nonlinearity is essential for bosonic reservoirs that use continuous variables as output features, in implementations such as ours that use discrete-variable outputs, the Kerr effect adds to the measurement-induced nonlinearity. To investigate its influence, we performed simulations isolating the effect of Kerr nonlinearity on classification performance. Our results show that Kerr nonlinearity is beneficial, especially in regimes where the input amplitudes interact non-trivially with the detuned resonator dynamics. These findings identify Kerr-induced nonlinearity as a useful computational resource, whose role in temporal and quantum tasks will be explored in future work.
Looking forward, this scheme can be scaled up by capacitively or parametrically coupling multiple resonators, each with a dedicated ancilla qubit. In such scaled architectures, the robustness of trained readouts to slow parameter drifts and environmental non-stationarities will become an increasingly important consideration. In the present work, all training and testing data are acquired within a single continuous measurement session, during which the system dynamics remain stationary and the trained readout remains valid. While this justifies the performance reported here, the long-term stability of trained reservoirs has not been systematically investigated. Quantifying drift timescales, sensitivity to small fluctuations, and the associated retraining requirements will be crucial for scalable circuit QED–based reservoir computing and constitutes an important direction for future work.

\begin{figure*}[t]
    \centering
    \includegraphics[width= 0.8\textwidth]{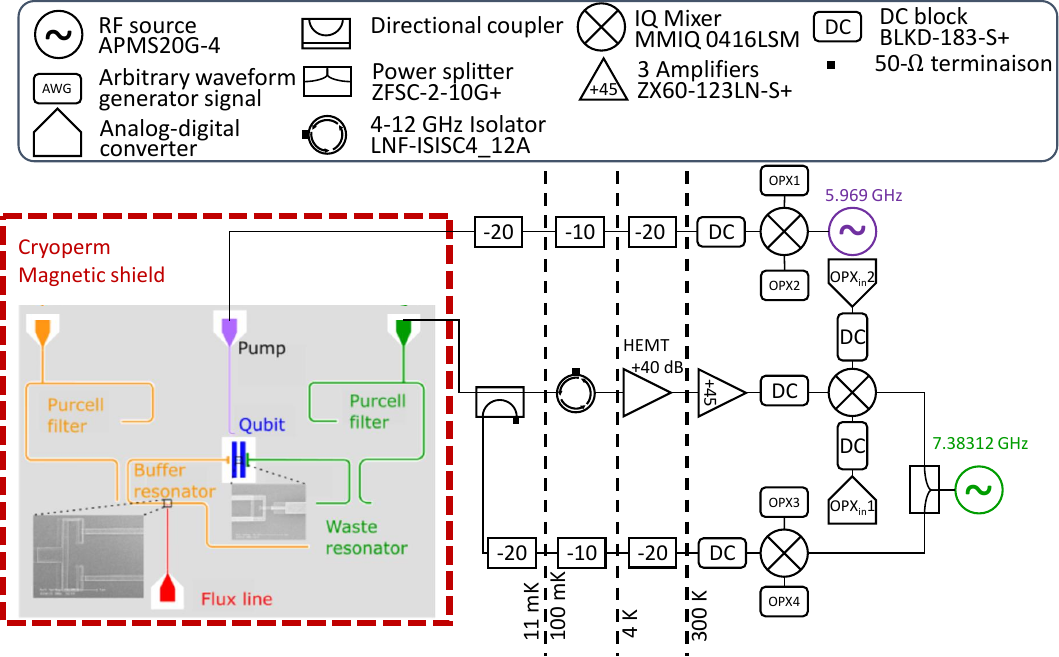}
    \caption{\label{FigureAFridge}
    \justifying 
    Schematic of the experimental setup at low and room temperature. The sample is placed on the base stage of a dilution cryostat at 11 mK, inside a cryoperm electromagnetic shield. The sample design is adapted from \cite{Balembois_thesis}. In our current setup, the flux noise was too strong, making the second resonator (yellow) unusable in practice.
    }
    
\end{figure*}

\section*{Acknowledgments}

The authors thank E. Flurin (SPEC, CEA, Gif-sur-Yvette, France) for his help in the sample fabrication and advice on performing the circuit measurements. This research was supported by the Paris Ile-de-France Region in the framework of DIM SIRTEQ and by European Union (ERC, qDynnet, 101076898).

\section*{Data and code availability}

The data and code that support this study are available in Ref.~\cite{carles2025data}.

\section{Appendix: Device and measurement setup}
\subsection{Device}

The device is a single microwave photon detector shown in Fig. \ref{FigureAFridge} fabricated by L. Balembois and E. Flurin in SPEC, CEA~\cite{Balembois_thesis}. In the present work, we use the waste resonator (green) as the cavity. The relevant circuit parameters are summarized in Table~\ref{T_circuit_param}.

\begin{table}[ht]
\centering
\begin{tabular}{|l|l|l|}
\hline
Circuit parameter & Symbol & Value \\
\hline
Cavity frequency & $\omega_c/2\pi$ & 7.617 GHz \\
Qubit frequency & $\omega_q/2\pi$ & 6.21031 GHz \\
Cavity-qubit cross Kerr rate & $\chi/2\pi$ & 22.29 MHz \\
Qubit decay time & $T_1$ & 8.01~$\mu$s \\
Cavity decay time & $T_c$ & 0.93~$\mu$s \\
\hline
\end{tabular}
\caption{Circuit parameters.}
\label{T_circuit_param}
\end{table}

\subsection{Measurement setup}

The experimental setup is shown in Fig.~\ref{FigureAFridge}. The cavity and qubit drives are generated by an advanced quantum control and processing unit, the OPX+ by Quantum Machines, which comprises an arbitrary waveform generator (AWG) and a fast acquisition card. Pulses are modulated at 233.88 MHz for the cavity and at 241.31, 219.3, 198.0, 177.1, and 157.4 MHz for the qubit, with a sampling rate of 1 GHz. I–Q mixers are used to combine these RF pulses with local-oscillator tones generated by an Anapico APMS20G at 7.38312 and 5.969 GHz.

The reflected signal is amplified by a HEMT amplifier at 4 K and by three 15 dB amplifiers at room temperature. The signal is then down-converted using I–Q mixers and digitized by the OPX+. DC blocks are added on the input lines to maintain a 50~$\Omega$ impedance match, as well as after amplification to improve digitization. The OPX+ further amplifies the output signal by 20 dB, digitizes it at a 1 GHz sampling rate, and performs the demodulation.

\section{Appendix: Calibration}

\begin{figure}[h!]
    \centering
    \includegraphics[width=\linewidth]{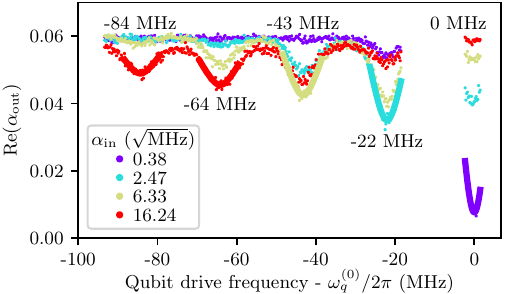}
    \caption{\label{FigureA1} \justifying
    Amplitude of the cavity output field at the end of the pulse sequence as a function of the qubit drive frequency for different color-coded cavity displacement amplitudes. The pulse sequence is composed of a cavity displacement pulse, followed by a 200~ns qubit pulse and a high cavity readout pulse. Full lines are fits centered around $\omega_q^{n}$ for the amplitude that yields $n$ photons in the cavity with highest probability.
    } 
\end{figure}

The cavity frequency $\omega_c$ and decay time $T_c$ are determined by fitting the cavity reflection coefficient. The occupation probabilities of the cavity are measured using qubit $\pi$ pulses conditioned on the cavity Fock state $|n\rangle$, denoted $\pi_{|n\rangle}$. The $\pi_{|0\rangle}$ pulse is implemented using a Gaussian pulse with a duration of 200 ns, corresponding to approximately $5/\chi$. Its frequency and amplitude are determined by fitting Rabi fringe measurements. All the other $\pi_{|n\rangle}$ pulses have the same shape, amplitude, and duration, while their frequencies are extracted from the two-tone spectroscopy shown in Fig.~\ref{FigureA1}. Specifically, the cavity output field amplitude $\alpha_\mathrm{out}$ is measured for different cavity displacements $\alpha_\mathrm{in}$ as a function of the frequency of a subsequent 200 ns qubit pulse. Dressed qubit resonance frequencies are extracted from the fitted minima of the cavity output field amplitude. The distance between different resonances corresponds to the cavity–qubit cross-Kerr rate $\chi$.

\begin{figure}[h!]
    \includegraphics[width=0.8\linewidth]{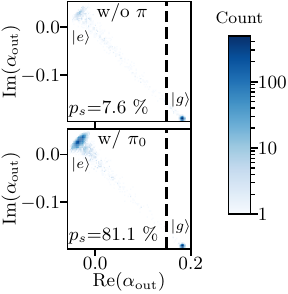}
    \caption{\label{FigureA2}\justifying Integrated cavity output signal on 10000 experiments represented in the quadrature space without the $\pi_{|0\rangle}$ pulse (top) and after a conditional $\pi_{|0\rangle}$ pulse on the qubit (bottom). The two clusters correspond to cavity states dressed by a qubit in the excited state (left cluster) or ground state (right cluster). The probability to ionize the qubit and thus to switch the cavity state corresponds to the proportion of points below the threshold $\mathrm{Re}(\alpha_{{out}})~=~0.15$ (dashed black line). 
    }
\end{figure}

\begin{figure}
\includegraphics[width=1\linewidth]{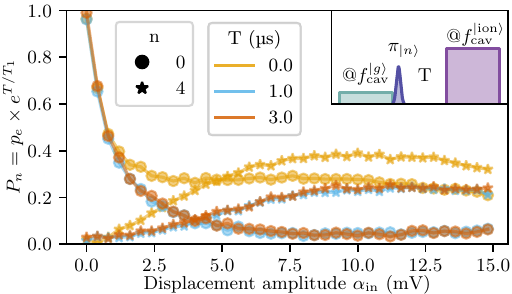}      
        \caption{\label{FigureAppendix3}      
               \justifying Occupation probabilities of the cavity states $|0\rangle$ (circles) and $|4\rangle$ (stars) as a function of the cavity displacement amplitude, for different color coded delays between the conditional $\pi_{|n\rangle}$ pulse and the high-power readout (HPR) pulse. Probabilities saturate for delays longer than 1~$\mu$s and are overestimated for shorter delays.
Inset: Pulse sequence associated with the measurement. The cavity displacement pulse (green) is sent at the cavity resonant frequency for the qubit in the ground state, $f^{|g\rangle}\mathrm{cav}$. The $\pi$ pulse conditioned on the cavity Fock state $|n\rangle$ is applied to the qubit (dark purple). Finally, the cavity is driven by the high-power readout pulse at the cavity resonant frequency for the ionized qubit state, $f^{|\mathrm{ion}\rangle}\mathrm{cav}$.
        }
\end{figure}

Qubit state measurement is performed using the high-power readout (HPR) as described in \cite{reed_high-fidelity_2010}. Figure~\ref{FigureA2} illustrates this technique by showing that the cavity response to a high-power drive at its bare frequency results in two distinct states in the quadrature space. These two states correspond to the cavity dressed by an ionized or non-ionized qubit and can be separated by applying a threshold at $\mathrm{Re}(a_\mathrm{out}) = 0.15$ (dashed line). Indeed, high-power cavity drive can lead to qubit ionization and thus a switch of the cavity resonant frequency. Since the cavity bare frequency corresponds to the resonance frequency of the cavity dressed by the ionized qubit, probing at this frequency results in a switch from a cavity dark state to a cavity bright state when the qubit becomes ionized. The readout amplitude is a key parameter of this measurement: if it is too low, the qubit will not be ionized even if it is initially in the excited state, whereas if it is too high, the qubit will be ionized even when it is in the ground state. The readout amplitude is therefore chosen to maximize the contrast between the switching probabilities with and without a $\pi_{|0\rangle}$ pulse. 

Furthermore, since the same cavity is used both for state preparation and readout, it is necessary to wait for cavity depletion before applying the HPR pulse. Otherwise, the effective readout amplitude would be increased by residual photons in the cavity, leading to unwanted qubit ionization and an overestimation of the switching probability. To identify the necessary delay time, we measure the occupation probabilities of states $|0\rangle$ and $|4\rangle$ for three different delays (0, 1 and 3 $\mu$s), shown in Fig.~\ref{FigureAppendix3}. The occupation probabilities $P_n$ are obtained by correcting for imperfections in the HPR measurement to extract $p_e$, and then multiplying by $e^{T/T_q}$, where $T_q$ is the qubit lifetime, to compensate for qubit decay during the delay. We observe that the probabilities saturate for delays longer than 1~µs and thus adopt the 1~µs delay between the $\pi_{|n\rangle}$ pulse and the readout for all the measurements. The reflected signal is recorded and demodulated to extract its envelope $\alpha_\mathrm{out}$, and a threshold $\mathrm{Re}(\alpha_\mathrm{out}) = 0.15$ mV is applied to determine the output state.

\bibliography{ExpQRC}

\end{document}